\documentstyle[aps,preprint]{revtex}

\begin{document}
\title{Risk aversion in financial decisions: A nonextensive approach\footnote
{{ Based on an invited conference given by one of us (C.T.) at the 
"International Public Seminar of the Year", 27 August 2002, Jakarta, Indonesia}} }

\author{C. Anteneodo and C. Tsallis\thanks{{\rm e-mail: celia@cbpf.br,tsallis@cbpf.br   }}  }

\address{Centro Brasileiro de Pesquisas F\'{\i}sicas, 
         R. Dr. Xavier Sigaud 150, \\
         22290-180, Rio de Janeiro, RJ, Brazil.  \\
         }

\maketitle

\begin{abstract}

The sensitivity to risk that most people (hence, financial operators) feel  
affects the dynamics of financial transactions.
Here we present an  approach to this problem based on a current generalization of 
Boltzmann-Gibbs statistical mechanics.

\end{abstract}

\vspace{1cm}


An important question in the theory of financial decisions is how to 
take into account those psychological attitudes of human beings that 
produce significant deviations from the ideally rational behavior. 
It is not by chance that a new discipline that focus on such questions, 
behavioral finance, is starting to gain universal recognition.  
In fact, Daniel Kanheman from the Psychology Department at Princeton University 
has been awarded (together with Vernon Smith) 
the 2002 Nobel Prize in Economics ``for having integrated insights 
from psychological research into economic science, 
especially concerning human judgment and decision-making under 
uncertainty'' \cite{nobel}.

Indeed, one of the human attitudes with important 
consequences in financial decision making is the  
risk aversion (attraction) that most people feel when they  
expect to gain (lose).
This sensitivity to risk is also observed in animals such as rats,
birds and honeybees \cite{animals} when they are exposed to 
variable food sources with different statistical properties, such 
as mean or variance, of the  offered quantity of  food.

The usual preference for a sure choice over an alternative of equally
or even more favorable expected value is called {\it risk aversion}.
Actually, most people present the tendency to feel aversion to risk 
when they expect to gain with moderate or high probability, and attraction
to risk when they expect to lose. 
However, these tendencies are inverted for very low probabilities\cite{at1}.

Naturally, this pattern of attitudes affects most human decisions 
since chance factors are always present, e.g., in medical strategies,
in gambling or in financial transactions.
In particular, in the context of finances,
the attitude of economic operators under risky choices
clearly is one of the main ingredients to be kept in mind
for realistically modeling market dynamics.


In the present text, we want to discuss the  sensitivity to risk within 
the context of nonextensive statistical mechanics \cite{ct1,further}. 
In order to do so we apply 
methods of statistical physics, a strategy that has proved to be very useful 
in several previous works \cite{physics} (see also \cite{econophys} for 
general discussions on the application of statistical physics methods in economics). 
The nonextensive formalism was introduced over a decade ago by 
one of us\cite{ct1} and further developed\cite{further}, with the aim 
of  extending the domain of applicability of statistical mechanics  
procedures to systems where Boltzmann-Gibbs (BG) standard formalism 
presents serious mathematical difficulties or just fails. Indeed, there is an  
increasing number of systems for which the standard mathematical expressions
of BG statistics appear to be inappropriate. Some of these cases can 
be satisfactorily treated within the new, nonextensive formalism. 
Therefore, a considerable amount of applications in many fields have 
been advanced in the literature \cite{applications}. 
The wide range of applications probably is deeply related to the  
ubiquity of fractal structures, power-laws, self-organized criticality in  nature.

The nonextensive statistics is based on the following entropic form

\begin{equation} \label{Sq}
S_q\;=\;k\frac{1-{\displaystyle \sum_{i=1}^W p_i^q} }{q-1} 
\;\;\;\;\;\;\;\;\;\;\;\mbox{with $\;\;\;\;{\displaystyle \sum_{i=1}^W p_i =1}$;  $\;\;\;q\in\Re$},
\end{equation}
where $W$ is the total number of microscopic configurations $i$ with probability  
$p_i$. This expression recovers, in the limit $q\to 1$, the usual
Boltzmann-Gibbs-Shannon entropic form

\begin{equation}
S_1\;=\;-k \sum_{i=1}^W p_i \, \ln p_i\, .
\end{equation}

Within the nonextensive formalism, suitable expectation values of a given 
quantity $A$ are calculated as {\em  normalized q-expectation values}, 
defined through

\begin{equation} \label{normev}
\langle\langle A \rangle\rangle_q \;\equiv\;
\frac{\displaystyle \sum_{i=1}^{W}p_i^q A_i}{\displaystyle \sum_{i=1}^{W}p_i^q},
\end{equation}
where $A_i$ is the value that the observable $A$  adopts in configuration $i$.


Coming back to economics, traditionally, the analysis of decision making under
risk was treated through the ``expected utility theory'' (EUT) \cite{eut},
on the assumption that individuals make rational choices.
More precisely, the {\it expected value} $E$, corresponding to the {\it 
prospect} ${\cal P}\equiv(x_1,p_1;\ldots;x_n,p_n)$
such that the outcome $x_i$ (gain if positive; loss if negative) 
occurs with probability $p_i$, is given by
$ E({\cal P}) \,=\,{\displaystyle \sum_{i=1}^n \chi(x_i)}\,p_i$,
where the weighting function $\chi(x_i)$ monotonically increases with $x_i$.
(Clearly, a statistically fair game corresponds to $\chi(x_i)=x_i$.) 
There are however aspects of risk sensitivity that are not adequately
contemplated within EUT. Such features were exhibited,
through experiments with hypothetical choice problems, by
Kahneman and Tversky\cite{at1}. They then proposed a generalization to EUT
equation within ``prospect theory'' (PT) \cite{at1}:  
$ E({\cal P}) \,=\,{\displaystyle \sum_{i=1}^n }\chi(x_i)\,\Pi(p_i)$, 
where the weighting function $\Pi(p_i)$ monotonically increases with $p_i$.

More recently, PT was generalized \cite{cpt} using a rank dependent 
or cumulative representation where the ``decision weight" 
multiplying the value of each outcome is distinguished from the 
probability weight. This interesting generalization is however 
irrelevant for the present discussion, where we will deal with 
simple prospects with a single positive outcome in which case both 
versions coincide.

The typical shape (corresponding to the most frequent human attitude)  
of the weight $\Pi(p_i)$ basically is, as sketched by Tversky and collaborators 
\cite{at1,weights} on the ground of experiments and observations, 
an increasing function,  concave for low  and convex for high 
probabilities,  with $\Pi(0)=0$, $\Pi(1)=1$ and $\Pi(p^*)=p^*$, for some $p^*$ typically verifying $0<p^*<1/2$.  
The following functional forms have been proposed \cite{ct1,applications} in the context 
of nonextensive statistical mechanics:
$ \Pi(p)= p^q,\;(q \in \Re)$ and 
$ \Pi(p)= p^q/( p^q + (1-p)^q ) $, 
usually referred to as {\it escort probability}. 
Other functional forms are also available in the literature \cite{shape}, 
such as
$ \Pi(p)= p^q / [p^q + (1-p)^q]^{1/q}$ and 
$ \Pi(p)= p^q / [p^q+ A(1-p)^q] $,  
where $A>0$. Clearly, $A=1$ recovers the escort probability. 
In all these cases, each individual can be characterized by a set of 
parameters which yields a particular $\Pi(p)$ representing
the subjective processing that a given individual makes of
known probabilities $p$  in a chance game.

In  the  regime of moderate and high probabilities, human behavior 
can be satisfactorily described by the weighting function $\Pi(p)=p^q$. 
This expression, which has a simpler form than other weights describing the full 
domain, is the one that we will adopt  throughout the present text.

Let us illustrate, through a simple example, the kind of choice problems we 
are referring to. The proponent of a transaction typically asks:
``What do you prefer: to receive with certainty \$ 85,000  
or to play a game where you receive \$100,000 with probability 0.85  
and nothing with probability 0.15?''. 
The game occurs only  once. In this case most people choose to take  the money.

Clearly, the present games are not the kind of operations that actually occur 
in a financial market. However in the sense of the theory of financial decisions, 
they paradigmatically illustrate the risk aversion phenomenon.

One can think in terms of normalized $q$-expectation values as follows

\begin{equation}
\langle\langle \mbox{gain}/ \mbox{take the money} \rangle\rangle_1 \;=\;85,000
\end{equation}
and

\begin{equation}
\langle\langle \mbox{gain}/ \mbox{play the game} \rangle\rangle_q \;=\;
\frac{100,000\times0.85^q+0\times0.15^q}{0.85^q+0.15^q}
\end{equation}

Notice that the standard expectation value of the game is also \$ 85,000; this corresponds to an ideally rational player, i.e., $q=1$. 
Since most people prefer to take the money, this means that most people have $q<1$ 
for this particular decision game. For the loss problem, an analogous reasoning 
leads to $q<1$ also, therefore unifying both situations.

Now, how can we measure the value of $q$ that characterizes the attitude of 
an individual in connection with a particular game?
The person is asked to choose between having the quantity $X$ in hands or 
playing the game of receiving $Y=$\$100,000 with probability $P=0.85$ 
and nothing with probability $0.15$. Then we keep changing (typically decreasing) the value of $X$ and
asking again until the person changes his(her) mind at a certain value $X_c$. 
Then, the value of $q$ associated with that person, for that problem, is given by
the equality

\begin{equation}
V_c\;=\;
\frac{100,000\times0.85^q+0\times0.15^q}{0.85^q+0.15^q}
\end{equation}
In particular if the threshold value is $85,000$, this means that the individual 
acts rationally, with $q=1$.

If unnormalized q-expectation values were considered instead of (\ref{normev}), i.e., 
if $ \langle\langle A \rangle\rangle_q \;\equiv\;\sum_{i=1}^{W}p_i^q A_i$,
then it is easy to show that most individuals act with $q>1$.

In a recent  work\cite{model} we investigated the consequences of risk averse 
attitudes in the dynamics of economic operations. 
We introduced an automaton simulating monetary transactions among 
operators with different attitudes under risky choices.  
Elementary operations were of the standard type used in hypothetical choice 
problems that exhibit risk aversion \cite{at1}, that is, of the type illustrated above. 
By following  the time evolution of the asset position of the operators, it is
possible to conclude on the consequences of each particular attitude.
We concentrated on problems where moderate or high probabilities are involved.

We considered different cases:
in A (alter-referential), the proponent operator somehow knows
the psychology of the other (characterized by $q'$);
in S (self-referential), the proponent ignores $q'$ and 
attributes to the other operator his/her own value of $q$; finally,
in C (consensual), the two operators act by consensus. 
Different restriction rules on the level of indebtedness of the operators were also
considered in the model.

One observes that the type of conditions limiting indebtedness are critical 
for defining the nature of the long term evolution, i.e., existence or not of
a nontrivial steady state. 
If individuals become {\em permanently} forbidden to trade from the 
instant their assets become less than a minimal quantity $M^*$ 
(restrictions of type PR, standing for {\it permanent restraints}) then the assets evolve to a trivial steady state
where there is concentration of wealth around the more rational player 
(a Dirac $\delta$-function centered at $q=1$ or at the boundary closer to 
$q=1$). This result is independent of the initial distribution of $q$.

We also considered {\em opportunistic} indebtment restraints (type OR, standing for {\it opportunistic restraints}) 
where agents can operate indefinitely except that they 
do not pay when they would have to do so if at a given step of the dynamics 
their assets become less the minimal quantity $M^*$ (i.e., operators can 
become swindlers occasionally).
In this case the system evolves to a nontrivial steady state. 
The details of this steady state depend,  
among other factors, on the distribution of the parameter $q$ of the operators. 
In Fig. 1 (a), we exhibit the average amount of money of the operators $\bar M(q,t)$ 
as a function of their $q$ for different time instants   
(the average is taken over a large number of realizations (histories) ). 
The initial distribution of $q$ was a uniform distribution in $[0,4]$ 
since about 75$\%$ of the people are risk-averse when high probabilities are involved 
(in the simulations we considered unnormalized expected values, 
therefore most individuals act with $q>1$). 
The maximum of the distribution depends on the hypothesis made on the value of 
$q$ of the partner. For a hypothesis of type A, the rational player wins, for
type S there are maxima on both sides of $q=1$ (the absolute one being for 
$q>1$, i.e, agents who are conservative for gains). For the consensus case C, 
the maximum asset occurs for $q>1$  (for more details see Ref. \cite{model})

Interestingly enough, some level of tolerance with regard to those who owe money 
avoids extreme wealth inequality to become the stationary state.  
However, one must keep in mind that in our simulations the
distribution of $q$ is kept fixed along the dynamics and, therefore,
the psychological effect of asset position is not being taken into account
in the present model. The inclusion of such ingredient in the dynamics would provide an improved, 
more realistic model. 
\\[1cm] \noindent
{\bf Acknowledgements:} 
One of us (C.T.) acknowledges warm hospitality at the Conference in Jakarta. 
We acknowledge Brazilian agencies CNPq, FAPERJ and PRONEX/MCT for financial support.

\begin{figure}[htb] 
\unitlength 1mm 
\caption{\protect 
Time evolution of assets with indebtedness restraint of kind OR 
(without exclusion of those who are indebted) with threshold $M^*=100$.
(a) $\bar{M}(q,t)-M_o)/M_o$ {\em vs.} $q$ at term $t/N =25000$ when the steady state is already attained. 
Lines correspond to simulations averaged on $2\times10^3$ histories with 
uniform initial assets $M(q,0)=M_o=1000$, 
number of agents $N=40$, 
quota interchanged in the game $S=100$ and probability for playing the game $P=0.85$. 
(b) $\bar{M}(q_{max},t)$ {\em vs.} $t$ and
(c) $q_{max}$ {\em vs.} $t$, where $q_{max}$ maximizes $\bar{M}(q,t)$.
The initial distribution of assets is uniform in [0,4]. 
The steady state does not depend on the initial distribution of assets.
}

\end{figure}


\begin{thebibliography}{99}
\bibitem{nobel} http://www.almaz.com/nobel.
\bibitem{animals} Ito M., Takatsure S., Saeki D.,
{\it J. of Exp. Analysis of Behavior} {\bf 73}, 79 (2000);
Shafir S., Wiegmann D. D., Smith B. H. and Real L. A.,
{\it Animal behavior} {\bf 57}, 1055 (1999).


\bibitem{at1} Kahneman D. and Tversky A., {\it Econometrica} {\bf 47},  263 (1979).


\bibitem{ct1} Tsallis C., {\it J. Stat. Phys.} {\bf 52},  479 (1988).


\bibitem{further} Curado E.M.F. and Tsallis C.,  J. Phys. A {\bf 24}, L69 (1991); 
Corrigenda: 24, 3187 (1991) and 25, 1019 (1992); A 24, L69 (1991); 
Tsallis C., Mendes R. S. and Plastino A.R., Physica A {\bf 261}, 534 (1998); 
see http://tsallis.cat.cbpf.br/biblio.htm for an updated bibliography.


\bibitem{physics}  Borland L., {\it Phys. Rev. Lett} {\bf 89}, 098701 (2002); 
Bouchaud J.P. and Mezard M., {\it Physica A} {\bf 282}, 536 (2000);
Bouchaud J.P., Potters M. and  Meyer M., {\it Eur. Phys. J. B} {\bf 13}, 595 (2000);
Bouchaud J.P., Matacz A. and  Potters M., {\it Phys. Rev. Lett.} {\bf 87}, 8701 (2001);  
Ausloos M. and Ivanova K., {\it Eur. Phys. J. B} {\bf 20}, 537 (2001);  
Ausloos M., {\it Physica A} {\bf 285}, 48 (2000); 
Andersen J.V., Gluzman S. and Sornette D., {\it Eur. Phys. J. B} {\bf 14}, 579 (2000);  
Sornette D., {\it Physica A} {\bf 284}, 355 (2000);   
Johansen A. and Sornette D., {\it Eur. Phys. J. B} {\bf 17}, 319 (2000); 
Stanley H.E., Amaral L.A.N., Buldyrev S.V., Gopikrishnan P., Plerou V. and Salinger M.A.,
       {\it P. Natl. Acad. Sci. USA} {\bf 99}, 2561 (2002); 
Stanley H.E., Amaral L.A.N., Gopikrishnan P.,  Plerou V. and Salinger M.A.,
    {\it J. Phys. Cond. Mat} {\bf 14}, 2121 (2002);
Borges, E. P., preprint cond-mat/0205520.


\bibitem{econophys} Stanley H.E., Amaral L.A.N., Gabaix X., Gopikrishnan P. and Plerou V.
{\it Physica A} {\bf 299}, 1 (2001); Ausloos M., {\it Eur. Phys. J. B} {\bf 20}, U2 (2001). 


\bibitem{applications}
Salinas S.R.A.  and Tsallis C., eds., {\it Nonextensive Statistical Mechanics and 
Thermodynamics}, {\it Braz. J. Phys.} {\bf 29} (1999);
Abe S. and Okamoto Y., eds. {\it Nonextensive Statistical Mechanics and Its 
Applications}, Series {\it Lecture Notes in Physics} (Springer-Verlag, Heidelberg, 2001);
Grigolini P., Tsallis C. and West B., eds., {\it Classical and Quantum Complexity and Non-extensive Thermodynamics}, {\it Chaos, Solitons and Fractals} {\bf 13} (3) (2002);
Kaniadakis G., Lissia M., and Rapisarda A., eds., {\it Non Extensive Statistical Mechanics and 
Physical Applications}, Physica A {\bf 305} (Elsevier, Amsterdam, 2002); 
Gell-Mann M. and Tsallis C., eds., {\it Nonextensive Entropy - Interdisciplinary 
Applications}  (Oxford University Press, Oxford, 2003), in press; 
Swinney H.L. and Tsallis  C., eds., {\it Anomalous Distributions, Nonlinear Dynamics, and
Nonextensivity}, Physica D (Elsevier, Amsterdam, 2003), in press. 



\bibitem{eut} Fishburn P. C., {\it Utility theory for decision-making},
New York, Wiley (1970).


\bibitem{weights} Tversky A. and Wakker P., {\it Econometrica}, {\bf 63} (1995) 1255;
Tversky A. and Fox C.R., {\it Psychological Review} {\bf 102}, (1995) 269.


\bibitem{shape} Gonzalez R. and Wu G., {\it Cognitive Psychology } 
{\bf 38},  129 (1999).


\bibitem{cpt} Tversky A. and Kahneman D., 
{\it Journal of Risk and Uncertainty} {\bf 5},  297 (1992).


\bibitem{model} Anteneodo C., Tsallis C., and Martinez A.S., {\it Europhys. Lett.}  
{\bf 59}, 635 (2002).


\end{thebibliography}
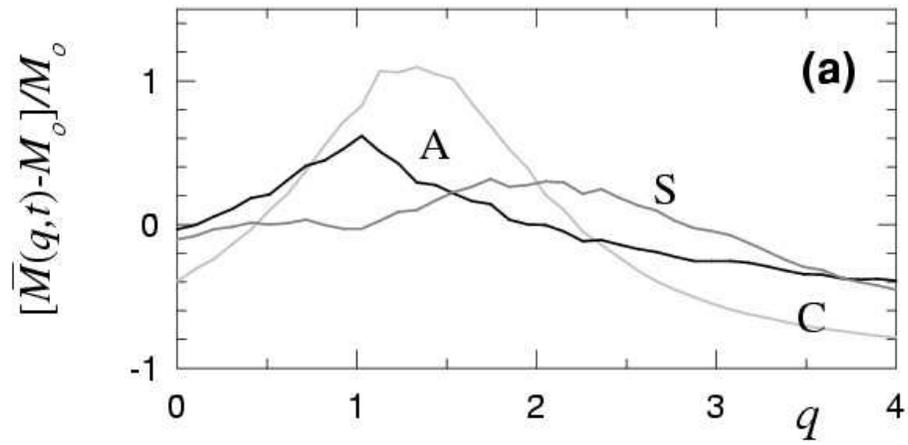
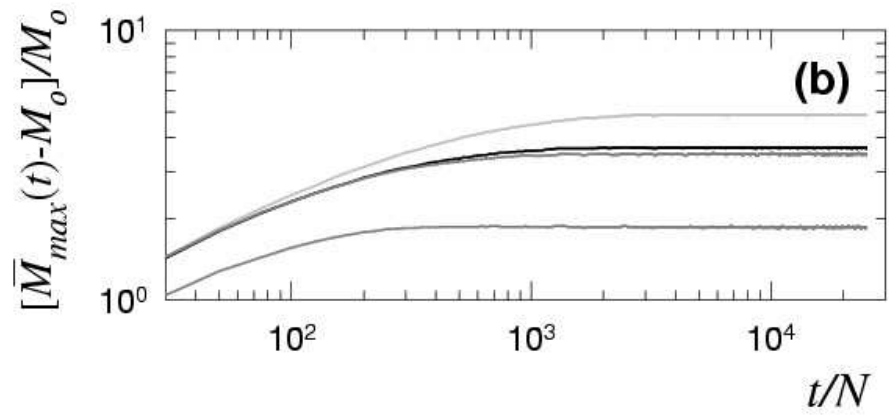
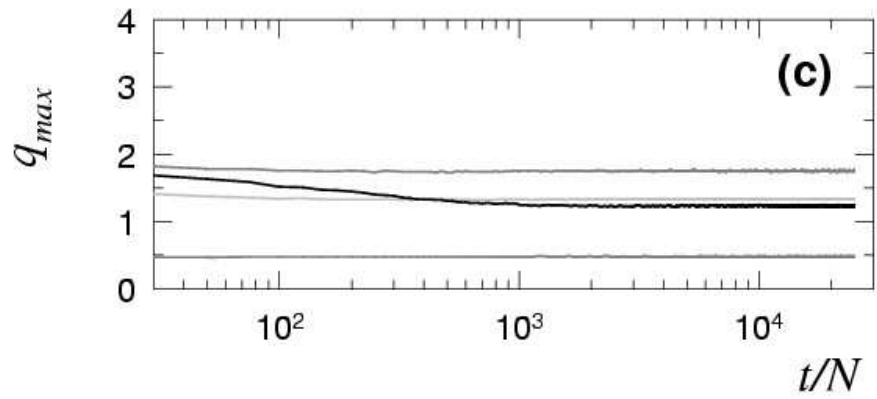
\end{document}